# Helium Ion Microscopy for Reduced Spin Orbit Torque Switching Currents

*Peter Dunne[1,2*‡], Ciaran Fowley[3*‡], Gregor Hlawacek[3], Jinu Kurian[1], Gwenael Atcheson[4], Silviu Colis[1], Niclas Teichert[4], Bohdan Kundys[1], Munuswamy Venkatesan[4], Jürgen Lindner[3], Alina Maria Deac[3], Thomas M. Hermans[2], J. M. D. Coey[4], Bernard Doudin[1]*

[1]Université de Strasbourg, CNRS, IPCMS UMR 7504, 23 rue du Loess, F-67034 Strasbourg, France

[2]Université de Strasbourg, CNRS, ISIS, 8 allée Gaspard Monge, 67000 Strasbourg, France

[3]Institute of Ion Beam Physics and Materials Research, Helmholtz-Zentrum Dresden - Rossendorf, Bautzner Landstraße 400, 01328 Dresden, Germany

[4]AMBER and School of Physics, Trinity College, Dublin 2, Ireland

ABSTRACT

Spin orbit torque driven switching is a favourable way to manipulate nanoscale magnetic objects for both memory and wireless communication devices. The critical current required to switch from one magnetic state to another depends on the geometry and the intrinsic properties of the materials used, which are difficult to control locally. Here we demonstrate how focused helium ion beam irradiation can modulate the local magnetic anisotropy of a Co thin film at the microscopic scale. Real-time *in-situ* characterisation using the anomalous Hall effect showed up to an order of magnitude reduction of the magnetic anisotropy under irradiation, and using this, multi-level switching is demonstrated. The result is that spin-switching current densities, down to 800 kA cm$^{-2}$, can be achieved on predetermined areas of the film, without the need for lithography. The ability to vary critical currents spatially has implications not only for storage elements, but also neuromorphic and probabilistic computing.

MAIN TEXT

Spin transport across material interfaces is sensitive to the electronic and structural nature of the interface. Since the discovery of giant magnetoresistance in the late 1980's[1,2] through to the theoretical[3] and experimental realisation of spin transfer torque[4,5], spintronics is a younger field that has grown steadily, with modern data storage already exploiting conventional spin transfer torque to change magnetisation states in magnetic tunnel junctions[6]. Spin orbit torque (SOT) switching, on the other hand, is a relatively young field[7–9], relying on the spin Hall[9] and Rashba[10] effects to manipulate static and dynamic magnetisation states by the flow of electrical current in adjacent heavy metal (HM) layers. In contrast to spin-transfer-torque MRAMs[11], SOT devices require less demanding 3D fabrication, greatly simplifying their production. Moreover, their planar nature allows for easy visualisation and easier exploitation of the stray magnetic fields in applications such as reprogrammable magnetic domain configurations for spin-wave logic[12].

The present drive to reduce switching currents and energy consumption is constrained by the selection of materials necessary to switch a magnetic domain *via* SOT. Similarly, real devices require multi-step nanofabrication to define memory storage cells. Here we present a new method to both reduce the switching currents, and to avoid multi-step lithography by using He$^+$ irradiation to locally modify the material properties in a single layer. Consequently, we can reduce the critical current density in a macroscopic device by almost an order of magnitude. Furthermore, we demonstrate how *in-situ* electrical measurement during irradiation allows precise control over the manipulation of the perpendicular magnetic anisotropy, capturing its evolution as a function of ion dose, making possible high-precision control and reproducibility of the process.

Light ion irradiation can lead to structural reorganization of stable material phases while preserving overall atomic composition with little to no topological damage[13]. This allows for the reduction of PMA in thin magnetic multilayers[14,15], which is caused by the intermixing of layers and increase of interfacial roughness[13]. Furthermore, He$^+$ and Ar$^+$ broad-beam irradiation has been used to tailor SOT driven domain-wall dynamics[16], improve the spectral linewidth of SOT nano-oscillators[17], and reduce the critical current density to switch the magnetisation of extended thin films via SOT[18]. Patterned magnetic anisotropy using broad beam irradiation has been demonstrated using shadow masks, but the smallest patterned volumes are mostly determined by the lithography process rather than the ion-beam interaction volume[15]. On the contrary, focused helium ion beams allow precise control of perpendicular magnetic anisotropy[19,20] and magnetisation[21] on the order of size of the collision cascade (< ~10 nm)[22]. With helium ion microscope (HIM) based He[+] irradiation, the lateral dimensions of the magnetic structures are limited only by the collision cascade[22] due to the minimal beam-size of only 0.5 nm.

Samples of Ta(5)/Pt(2)/Co(1.0)/W(1.5)/Pt(1.5)/Ta(1.5) were prepared by magnetron sputtering on Si wafers, and patterned into 10 µm Hall bar structures by a combination of UV lithography, ion milling and lift-off. An asymmetric stack of Pt, Co, and W was chosen, as the opposite spin Hall angles of the Pt-Co and Co-W interfaces are known to maximise SOT switching efficiency[23]. SQUID magnetometry on unpatterned films with the substrate orientated parallel and perpendicular to the applied field indicated robust PMA (Fig. 1a), as expected for a sub-nm Co film[24], with an effective anisotropy field of 1.28 T, and a saturation magnetisation $\mu_0 M_s$ of 1.02 T. Anomalous Hall effect measurements confirmed that the PMA remains intact on the patterned Hall bars (Figs. 1b,c). Here, we found square hysteresis loops and coercivities of ~20 mT (Fig. 1b) for magnetic fields applied perpendicular to the film along *z*, and effective magnetic anisotropies of well over 1 T (Fig. 1c) for hard axis measurements (along *x*). At positive saturation (+$m_z$), the anomalous Hall resistance is 0.59 Ω.

Current-induced ($I_x$) switching measurements under a bias field $\mu_0 H_x$ = 150 mT (Fig 1d) yield a critical current density, $J_c$, of 6.0 MA cm$^{-2}$ for full magnetisation reversal in the microfabricated structure. To minimise thermal effects while sweeping the current, $I_x$ was pulsed for a duration of 2 ms, with a delay of 1 s between subsequent pulses, i.e., a duty cycle of 0.2%. Similar to the field driven switching in Fig. 1b, there is only one step and the saturated state has a resistance of 0.57 Ω.

The patterned devices were locally irradiated with helium ions with a 30 kV acceleration voltage using an Orion Nanofab helium ion microscope system, while in real time, we monitored the evolution of the anomalous Hall resistance *in-situ*. This electrical characterisation system consists of four Kleindieck MM3A micromanipulators which allow three-dimensional positioning of the contacting needles in a high vacuum environment (see Fig. 2a). A Keithley 2400 source meter was used to apply current and read the voltage drop corresponding to the longitudinal and transverse resistances. A current of 1 mA applied along the *x*-axis is used to probe the evolution of the anomalous Hall resistance close to zero applied magnetic field. The combination of the HIM with the possibility for *in-situ* control of the magnetisation state via Hall resistance measurements is unique in that it allows unprecedented spatial resolution, high flexibility in the patterning by avoiding high resolution but low yield lithography steps, resulting in a high turn around for explorative experiments. This compliments alternative approaches based on laser illumination with a 1.5 µm lateral resolution[25], ion broad beam irradiation which requires thick photoresist masks that limit the achievable spatial resolution[18], electric field control of anisotropy[26], or even oxidation which is CMOS compatible but inflexible and not ideal for fundamental studies[27].

During irradiation, the anomalous Hall resistance steadily decreases (Fig. 2b) until a critical dose of 35 ions nm$^{-2}$ is reached, and a sharp switching is observed. The sharp fall in $\Delta R$ is complete at 37 ions nm$^{-2}$. The total fall of resistance (0.52 Ω) is close to the full deflection observed in the *ex-situ* measurements shown in Fig. 1, marked as a grey dashed line, and indicates a reduced effective anisotropy field in the irradiated zone ($H_{K2}$) compared to that of the extended structure ($H_{K1}$) (Fig. 2b). The layer intermixing introduced by irradiation gives control over $K_{eff}$, and $H_{K2}$ thus becomes an experimentally tuneable parameter. Although the irradiations were carried out over the complete area of the Hall cross, as shown in Fig. 2c, the close proximity of the unirradiated areas allow for residual information from those regions to be gathered in the Hall loops, leading to the offset of ~0.1 Ω in Fig. 2b.

We selected four doses along this irradiation curve to demonstrate the ease of applying this technique to the reduction of critical currents for spin orbit torque switching. Figure 2d shows the corresponding anomalous Hall response versus $\mu_0 H_x$ for 1, 20, 30, and 50 ions nm$^{-2}$. Aside from the reduction in magnetic anisotropy field, evidenced by the saturation of $R_{AHE}$ at lower in-plane applied fields, several steps are observed in the anomalous Hall loops for irradiation doses $\geq 30$ ions nm$^{-2}$. The additional steps correspond to the unchanged anisotropy in the unirradiated adjacent regions. This offset is highly dependent on the accuracy of overlaying the irradiation box with the junction area, and the residual Hall resistance is never precisely the same for each junction. However, the multi-step switching, as seen in Fig, 2d, is always a result of the reversal of neighbouring regions adjacent to the Hall cross.

Assuming coherent rotation of the magnetisation, the normalised first quadrant magnetisation, $M'_z$, (Fig. 2e) can be fitted with[28]

$$M'_z = \sqrt{1 - \left(\frac{\mu_0 H_x}{\mu_0 H_K}\right)^2} \qquad (1)$$

which yields the magnetic anisotropy field, $\mu_0 H_K$, as a function of irradiation dose (Fig. 2f). We observe a continuous decrease in the anisotropy field, from 1.32 T for as-deposited films, to only 0.12 T for samples irradiated beyond the critical dose of 35 ions nm$^{-2}$. The saturation magnetisation, $M_s$, (Fig. 2f) was determined from the relative change in the saturation Hall resistance with ion dose, when the magnetic field is applied out-of-plane, along $z$ (not shown here). $M_s$ is reduced at higher ion doses, due to alloying of the Co/HM interfaces, but is less sensitive than the magnetic anisotropy field, $H_K$, only reducing to 86% of the initial value compared to 33% for the anisotropy field after a dose of 30 ions nm$^{-2}$.

Both the $H_K$ and $M_s$ play an important role in current-induced SOT switching. Under a macrospin approximation with the condition $H_x \ll H_K$, where $H_x$ is a magnetic bias field applied in-plane, the critical current density, $J_c$, required to switch the magnetisation direction of a magnetic state (in SI units) is[29]

$$J_c = \frac{2e\mu_0 M_s t_f}{\hbar \theta_{SH}} \left( \frac{H_K}{2} - \frac{H_x}{\sqrt{2}} \right) \quad (2)$$

where e is the electronic charge, $\hbar$ is the reduced Planck constant, $t_f$ is the magnetic free-layer thickness, and $\theta_{SH}$ is the spin Hall angle. However, when the applied field is comparable to the anisotropy, as we find for increased ion doses, the critical current density can instead be written as[29]

$$J_c = \frac{2e\mu_0 M_s t_f}{\hbar \theta_{SH}} \left( \sqrt{\frac{H_K^2}{32} \left[ 8 + 20b^2 - b^4 - b(8 + b^2)^{\frac{3}{2}} \right]} \right) \quad (3)$$

where $b = \frac{H_x}{H_K}$. The effect of ion dose on $J_c$ is initially minimal for doses up to 20 ions nm$^{-2}$ (Figs. 2f, 3a, & Table 1), with $J_c^-$ reducing from -6.0 to -5.0 MA cm$^{-2}$ (for $\mu_0 H_x$ = 150 mT). For 30 ions nm$^{-2}$, close to the critical dose, the critical current density is, however, reduced by almost one order of magnitude to -800 kA cm$^{-2}$. At larger doses, 50 ions nm$^{-2}$, we only observe the switching in the peripheral regions adjacent to the Hall cross.

Nonetheless, reducing the critical current density is not the sole criterion for improvement of SOT switching of magnet states. Additionally, we define a SOT switching efficiency, $\eta$, being the ratio of the magnetic *energy* barrier ($K_{eff}$) per unit area (in the magnetic free layer $t_f$) versus the electrical *power* (P) per unit area (in the entire stack thickness d) required to switch the magnetic state, in units of seconds:

$$\eta = \frac{K_{eff} t_f}{Pd} = \frac{\mu_0 M_s (H_K - H_x) t_f}{2 J_c^2 \rho d} \quad (4)$$

where $\rho$ is the total resistivity of the device. The choice of units is deliberate; the minimum pulse length for deterministic SOT switching is not always reported, but if it is known, then $\eta$ can be divided by this duration to determine the dimensionless efficiency. Either way, a larger value of $\eta$ implies a more stable magnetic element that requires less power to switch. Defined in this way, the switching efficiency of different material systems can be compared regardless of the pulse length, so long as the bias fields are similar. Furthermore, increasing *N*, the number of Pt/Co/W repeats, will not impact the efficiency if the relative amount of magnetic material remains the same, i.e. $t_f/d$ is constant. If the bias fields are not similar, substituting $J_c$ from Eq. (2) into (4) under the condition $H_x \ll H_K$, leads to an intrinsic SOT efficiency for PMA systems:

$$\eta_i = \frac{\hbar^2 \theta_{SH}^2}{2\mu_0 M_s H_K e^2 \rho d t_f} \quad (5)$$

The advantage of Eq. (5) is that it depends solely on intrinsic material properties, and not on the experimental conditions, thus making it a more transparent comparator between different reported systems, particularly when the experimental current pulse may be much longer than the intrinsic switching time. Its drawback is that it is based on a macrospin approximation, which cannot envisage a multidomain response, and therefore overestimates the critical current density by a factor 10 –100.

Nevertheless, Eq. 5 provides a theoretical guideline for optimum parameters when considering material systems. This expression illustrates that to achieve large switching efficiencies, a minimal amount of conductive material must be used, while retaining a large spin Hall angle and a small anisotropy field. We find that neither the calculated spin Hall angle nor resistivity change significantly upon irradiation, with the former remaining ~0.4, and the latter ranging from 215 – 235 µΩ.cm. Similar to $J_c$, we find there is negligible change in the intrinsic switching efficiency, $\eta_i$, for irradiation doses < 20 ion nm$^{-2}$, which remains close to 2 fs (Table 1), increasing to a maximum of 4.2 fs at 30 ions nm$^{-2}$. In contrast, the experimental efficiency, $\eta$, is initially ~7.1 ps, and *drops* at 20 ions nm$^{-2}$ to 4.3 ps, before then increasing to its maximum of 86 ps at 30 ions nm$^{-2}$, an order of magnitude improvement upon the virgin state. The discrepancy between $\eta$ and $\eta_i$ is due to the breakdown of the macrospin approximation used to calculate the latter. In all cases, the switching efficiency compares favourably to the literature values listed in Table 2, where efficiencies are of order 0.01 – 10 ps, with a maximum of 69 ps found for metallic Ta/Pt/Co/W/Ta multilayers[30], and 21 ps for insulating barium ferrite on Pt[31], while zero-field switching devices[26], present much lower efficiencies (0.74 ps).

Returning to the SOT switching for 30 ions nm$^{-2}$, shown in Fig. 3a, we sketch the magnetic states at points 1, 2, 3 and 4 in Fig. 3b. As the interior of the Hall cross can be addressed separately from the rest of the device, this forms a two-state system that can be addressed electrically as shown in Fig. 3a, or magnetically as shown in Figs. 2c and 3c,d. Electrical switching involves applying a bias in-plane field (here 150 mT), and pulsing currents in a range of ± 3 mA to ensure switching of only the irradiated region. This leads to a switch between states 1 and 2, or 3 and 4 in (Fig. 3a,b). Otherwise pulsing currents of order ± 7.5 mA switches the magnetisation of the entire device between states 1 and 3 (Fig. 3a,b). Magneto-optic Kerr effect (MOKE) imaging under magnetic fields normal to the device ($\mu_0 H_z$) confirms that the entire irradiated region switches independently from the rest of the device, Fig. 3c. Individual hysteresis loops for the irradiated (30 ions nm$^{-2}$) and non-irradiated regions (as-dep$_1$, as-dep$_2$) are shown in Fig. 3d. The irradiated and unirradiated regions exhibit coercivities which are well spaced in magnetic field, $\mu_0 H_c$ = 5 mT and $\mu_0 H_c$ = 16 mT, respectively. The reader is directed to the supplementary video 1 to view the full magnetisation response during a field sweep.

Although there is an equivalence between electrical and magnetic switching, electrical SOT switching is more robust to variability in lithographic processing of the Hall bars than magnetic switching. This is because the former depends on the perpendicular magnetic anisotropy of a device, compared to coercivity for the latter. We find less than 1% variation in the magnetic anisotropy for the non-irradiated devices, $\mu_0 H_K$ =1.32 ± 0.01 T, as the anisotropy is determined by the deposition process, which in bulk, non-patterned, thin films is 1.28 ± 0.05 T. In contrast, we find a 26% variation of coercivity in the same devices, 14 ± 4 mT, reflecting the strong dependence of the coercivity on the lithographic processing and slight variations in dimensions of the Hall bars. Since we address the anisotropy directly with ion irradiation, this illustrates the advantageous precision and robustness of focussed ion beam irradiation for magnetic element patterning compared to material removal.

In order to assess the suitability of the device as a memory storage element, we analyse the effect of He$^+$ irradiation on the thermal stability of the created magnetic bits. As expected, the reduction in $J_c$ does come at the cost of thermal stability, defined by a parameter $\Delta = K_{eff} V / k_B T$, where $V$ is the magnetic element volume $k_B$ is Boltzmann's constant and $T$ is temperature. However, even for a 1 nm thick, 32 nm sided element, i.e. a 1024 nm$^3$ volume, $\Delta$ = 64 for doses as high as 20 ions nm$^{-2}$, more than the 10-year stability criterion of 60 used in magnetic recording[32]. Close to the critical dose, at 30 ions nm$^{-2}$, $\Delta$ drops to 38—the equivalent of 95% data retention for 16 days[33]— but unlike data storage, long-term stability is not critical to all applications. In fact, weakly bi-stable devices can be suitable for spin Hall oscillators, where the strength and modulation of spin wave propagation depends on the magnetic anisotropy[34]. An easy way to improve the thermal stability would be to add $N$ Pt/Co/W repeats to the system. $\Delta$ will increase linearly with $N$, while not impacting η (Eq. 4), as stated previously. Furthermore, the reduction in local magnetic anisotropy allows for minimal losses during magnetisation switching. This level of control highlights the effectiveness of this method to tailor the switching current density in PMA multilayers. Although only two steps are shown in Fig. 3, one can also write switching current spatial gradients of almost any value directly. This allows for complex tailoring of not only field free memory elements[35–38], but even permits the control of spin-wave waveguide locations coupled with on-and-off spin-wave transmission control[12,39]. Such switching current spatial gradients can be used to tailor the synaptic-like response from PMA memory elements[40] when applied to neuromorphic computing or even logic elements[27]. Extremely low thermal stability, would mean metastable 0 and 1 states which can be applied to probabilistic computing[41] or random number generation[42].

In summary, a He[+] microscope has been used to perform mask-less light-ion irradiation on perpendicular magnetic anisotropy stacks for local control of the SOT switching properties. *In-situ* electrical measurements of anisotropy reduction provides real-time control of the anisotropy, allowing accurate determination of the critical dose needed to reach a desired, up to the transition to in-plane magnetic anisotropy. The corresponding reduction in critical current densities for SOT switching are nearly linear with decreasing anisotropy. Irradiation at doses just below the critical dose allow us to achieve almost an order of magnitude control over the critical current required for robust switching, and DC switching current densities as low as 800 kA cm$^{-2}$ are achieved, which we believe is the lowest to date. Moreover, the switching efficiency of 86 ps is larger than the previous state of the art. It illustrates the unique advantages of this approach, and opens the door to preferential switching of predetermined areas of a device, limited only by the nm-resolution of the He$^+$ ion beam microscopy, while preserving the flat topography of the initial magnetic stack.

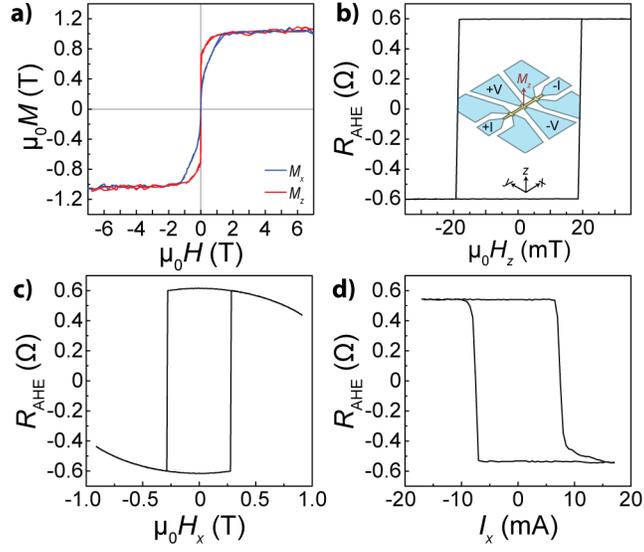

**Figure 1. a)** SQUID-VSM magnetometry parallel ($M_x$) and perpendicular ($M_z$) to the substrate. The anomalous Hall resistance of virgin junctions versus **b)** $\mu_0 H_z$; **c)** $\mu_0 H_x$; and **d)** applied DC pulse current, $I_x$, under a bias field of $\mu_0 H_x = 150$ mT.

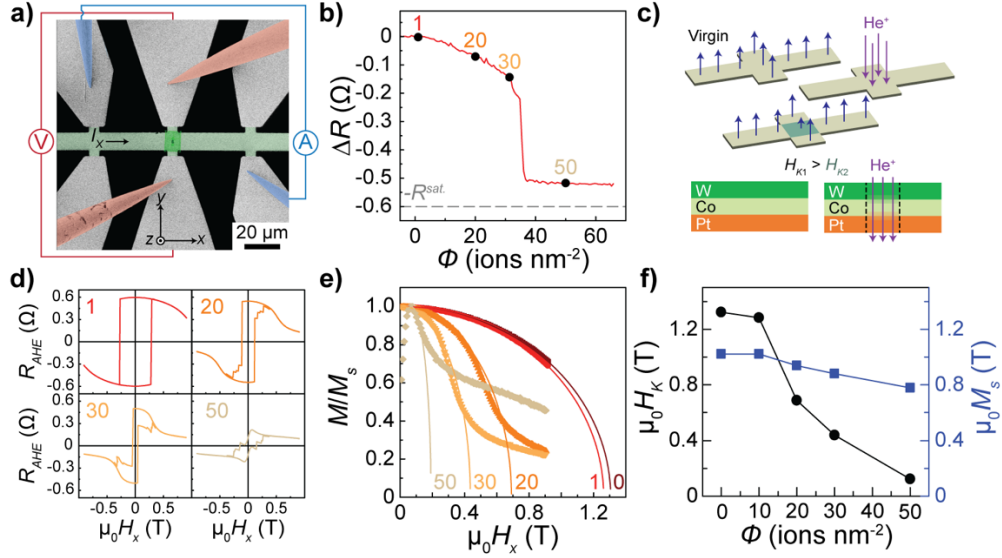

**Figure 2. a)** False coloured helium ion microscopy image of the *in-situ* contacting of a 10 μm Hall bar structure. The grey areas are the gold contact pads, the green stripe represents the magnetic multilayer under investigation, the red needle probes are for measuring the Hall voltage, while the blue ones are for applying the current. One such image corresponds to a dose of only 0.1 ions nm$^{-2}$. An overlay schematic shows the current and voltage lines used. **b)** *In-situ* change in anomalous Hall resistivity as a function of irradiation dose between 0 and 70 ions nm$^{-2}$. **c)** Schematic of the local irradiation process, whereby the local anisotropy is reduced due to interface mixing of the upper and lower Co / HM interfaces. **d)** *Ex-situ* $R_{AHE}$ versus $\mu_0 H_x$ curves at selected doses marked in a): 1 ion nm$^{-2}$, 20 ions nm$^{-2}$, 30 ions nm$^{-2}$, and 50 ions nm$^{-2}$. **e)** Normalised magnetisation from d) for all irradiated and virgin samples, with solid line fits using Eq. (1) to determine the anisotropy field, as a function of irradiation dose; **f)** Anisotropy field and saturation magnetisation determined from *ex-situ* anomalous Hall effect loops.

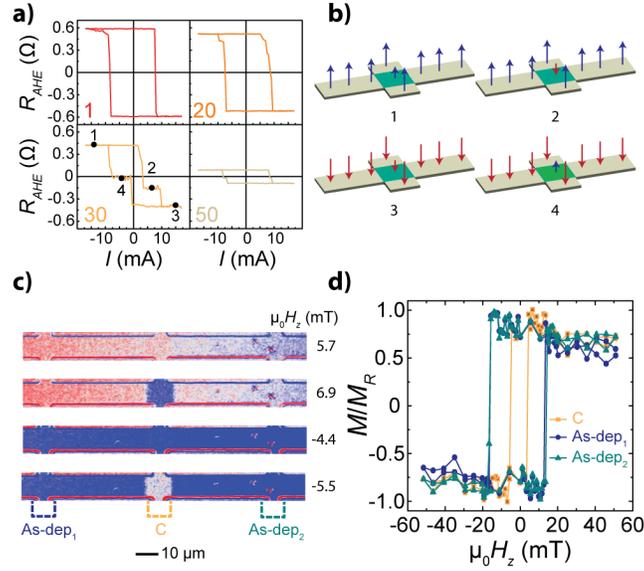

**Figure 3. a)** Current driven SOT switching curves at the same doses as Fig. 2 under a bias field of $\mu_0H_x = 150$ mT; **b)** Schematic of 4 distinct magnetic configurations achievable through electric current switching, numbered 1–4 in 30 ions nm$^{-2}$ sub-panel of panel a); **c)** MOKE images of the Hall bar under 4 different external fields, $\mu_0H_z$, displaying the 4 different magnetic states shown in panel b; **d)** MOKE-acquired hysteresis loops for the device receiving a dose of 30 ions nm$^{-2}$ at three different 10 × 10 μm$^2$ areas: one irradiated region and two non-irradiated regions as show in panel c).

**Table 1.** Irradiation dose dependency of the thin film magnetic properties.

| Dose ions nm$^{-2}$ | $K_{eff}$ kJ m$^{-3}$ | $\Delta$ | $\rho$ µΩ.cm | $I_c^-$ mA | $I_c^+$ mA | $J_c^-$ MA cm$^{-2}$ | $J_c^+$ MA cm$^{-2}$ | $J_c^{th}$ MA cm$^{-2}$ | $\eta$ ps | $\eta_i$ fs |
|---|---|---|---|---|---|---|---|---|---|---|
| 0  | 537 | 133 | 214 | -7.5 | 7.5 | -6.0 | 6.0 | 459 | 7.1 | 1.2 |
| 1  | 521 | 129 | 214 | -7.5 | 7.5 | -6.0 | 6.0 | 442 | 6.9 | 1.2 |
| 20 | 257 | 64  | 219 | -6.2 | 7.5 | -5.0 | 6.0 | 184 | 4.3 | 2.5 |
| 30 | 153 | 39  | 224 | -1.0 | 3.3 | -0.8 | 2.7 | 85  | 86  | 4.2 |
| 50 | 38  | 9   | 235 | –    | –   | –    | –   | –   | –   | –   |

The properties are: effective anisotropy $K_{eff}$, thermal stability parameter $\Delta$, negative and positive critical currents $I_c^-$, $I_c^+$, the corresponding experimental critical current densities $J_c^-$, $J_c^+$, the theoretical critical current density $J_c^{th}$ determined using Eq. (3) with $\mu_0 H_x$ = 150 mT, and an effective spin hall angle measured to be $\theta_{SH}$= 0.4, the switching efficiency $\eta$ calculated from Eq. (4), and intrinsic efficiency $\eta_i$ using Eq. (5). Note that for 50 ions nm$^{-2}$, only switching of adjacent unirradiated regions was detected.

**Table 2.** Comparing Switching Efficiencies.

| Stack | $J_c$ [a)] | $B_x$ | $\theta_{SH}$ [b)] | $\Delta$ [c)] | $P$ [d)] | $\eta$ | $\eta_i$ | Ref. |
|---|---|---|---|---|---|---|---|---|
| | MA cm$^{-2}$ | mT | | | PW m$^{-3}$ | ps | fs | |
| Pt 3/Co 0.6/AlO$_x$ 1.6 | 78 | 47.5 | 0.07 | 74 | 2330 | 0.02 | 0.09 | [10] |
| Ta 3/Pt 5/Co 0.6/Cr 2/Ta 5 | 2.70 | 20 | 0.41 | 106 | 0.89 | 28 | 2.02 | [43] |
| Ta 3/Pt 5/Co 0.8/W 3/TaO$_x$ 2 | 2.30 | 5 | 0.244 | 136 | 0.42 | 69 | 0.78 | [30] |
| Ta 5/CoFeB 1.1/MgO 2/Ta 2 | 3.09 | 20 | 0.08 | 56 | 1.91 | 14 | 0.19 | [44] |
| Bi$_2$Se$_3$ 7.4 [e)]/CoTb 4.6/SiN$_x$ 3 | 3.0 | 100 | 0.16 | 73 | 2.48 | 6.2 | 0.23 | [45] |
| Pt 5/BaFe$_{12}$O$_{19}$ 3 [f)] | 5.7 | 123 | 0.07 | 232 | 6.26 | 18 | 0.04 | [31] |
| Pt 4/Co 0.4/Ni 0.2/Co 0.4 /Pt 2 [g)] | 7.1 | 0 | 0.07 | 13 | 10.2 | 0.74 | 0.72 | [37] |

**a)** If $J_c^- \neq J_c^{+}$, then the smaller of the two values is taken; **b)** when not reported, literature values of $\theta_{SH}$ were taken from[46], and for multilayers with adjacent layers of opposite $\theta_{SH}$, the magnitude of the larger one was chosen; **c)** calculated for $\mu_0 H_x = 0$; **d)** when not reported, calculated using bulk resistivity values from[47], and then corrected using the M-S model[48]; **e)** topological insulator used for spin injection; **f)** ferromagnetic insulator; **g)** zero field switching device, moving between a remnant state $m_z = 0$, and a downward pointing $-m_z$ state.

## ASSOCIATED CONTENT
**Supporting Information**.

The following files are available free of charge.

Supporting video 1, "MOKE_C_30ions.avi":

*Ex-situ* normalised magnetisation of a 10 μm wide Hall bar under a magnetic field sweep from -50 mT to +50 mT and then back to -50mT using a MOKE microscope. The central 10 x 10 μm$^2$ area was irradiated with a dose of 30 ions/nm$^2$.

Supporting text, "DEPNA_single_layer_SI.pdf" consists of:

1. Derivation of equation (1) using a macrospin approximation, and an example fit to the sample irradiated with 20 ions nm$^{-2}$.

2. Anomalous Hall resistance plots for 8 non-irradiated samples for $\mu_0 H_z$ and $\mu_0 H_x$.

3. Discussion of the Mayadas–Shatzkes model used for applying thickness dependent corrections to bulk resistivities.

**Data and Code Availability** Source data for figures 1-3 and any other data that support the findings of this study, and the Python scripts used to process the data, are available on the Zenodo data repository: https://zenodo.org/communities/mami-h2020/

## AUTHOR INFORMATION

**Corresponding Authors**

*peter.dunne@ipcms.unistra.fr

*c.fowley@hzdr.de

**Author Contributions**

The manuscript was written through contributions of all authors. All authors have given approval to the final version of the manuscript.

**Funding Sources**

We acknowledge the support of the Labex NIE 11-LABX-0058_NIE within the Investissement d'Avenir program ANR-10-IDEX-0002-02. This project has received funding from the European Union's Horizon 2020 research and innovation programme under the Marie Skłodowska-Curie grant agreement No 766007.

**Notes**

The authors declare no competing financial interest.

**Acknowledgements**

We thank Fabien Chevrier for technical support, the staff of the STnano nanofabrication facility. The nanofabrication facilities (NanoFaRo) at the Ion Beam Center at the HZDR are also gratefully acknowledged.

**ABBREVIATIONS**

HM, Heavy metal; MRAM, magnetic random-access memory; PMA, perpendicular magnetic anisotropy; SOT, spin orbit torque

# Helium Ion Microscopy for Reduced Spin Orbit Torque Switching Currents: Supporting Information

Peter Dunne,[1,2,*] Ciaran Fowley,[3,†] Gregor Hlawacek,[3] Jinu Kurian,[1] Gwenael Atcheson,[4] Silviu Colis,[1] Niclas Teichert,[4] Bohdan Kundys,[1] Munuswamy Venkatesan,[4] Jürgen Lindner,[3] Alina Maria Deac,[3] Thomas M. Hermans,[2] J. M. D. Coey,[4] and Bernard Doudin[1]

[1]Université de Strasbourg, CNRS, IPCMS UMR 7504, 23 Rue du Loess, 67034 Strasbourg, France
[2]Université de Strasbourg, CNRS, ISIS, 8 allée Gaspard Monge, 67000 Strasbourg, France
[3]Institute of Ion Beam Physics and Materials Research,
Helmholtz–Zentrum Dresden – Rossendorf, Bautzner Landstraße 400, 01328 Dresden, Germany
[4]AMBER and School of Physics, Trinity College, Dublin 2, Ireland

## DETERMINING MAGNETIC ANISOTROPY FROM ANOMALOUS HALL EFFECT

If the magnetisation, **M**, of a perpendicularly magnetised thin film were to rotate coherently under an in-plane external magnetic field $H_x$, then its magnetic free energy can be described using the Stoner-Wohlfarth model as [1]

$$E = K_U \sin^2 \theta - \mu_0 M H_x \cos(\theta - \pi/2) \quad (1)$$

where $K_U$ is the uniaxial anisotropy, and $\theta$ is the angle between the magnetisation and the normal to the surface. Minimising the free energy with respect to $\theta$ results in

$$\theta = \sin^{-1}\left(\frac{\mu_0 M H_x}{2 K_U}\right) \quad (2)$$

The anomalous Hall effect is sensitive to the perpendicular component of the magnetisation, giving a signal proportional to $M_z$

$$M_z = M \cos \theta = M \cos\left[\sin^{-1}\left(\frac{\mu_0 M H_x}{2 K_u}\right)\right]$$
$$= M \sqrt{1 - \left(\frac{\mu_0 M H_x}{2 K_u}\right)^2} \quad (3)$$

For the normalised data used here, the equation is recast as

$$M'_z = \sqrt{1 - \left(\frac{B_x}{B_K}\right)^2} \quad (4)$$

where $M'_z$ is the normalised $z$ component of the magnetisation, $B_x = \mu_0 H_x$ and $B_K = \mu_0 H_K$ are the applied field and anisotropy field in units of T.
Normalised AHE data for a sample after irradiation (20 ions/nm$^2$) are shown in Fig. 1. The in-plane component, $M_x$ was assumed to be $M_x = \sqrt{1 - M_z^2}$, also shown in Fig. 1. The data was fitted using Eq. 4 from 0 to 300 mT, but above ≈430 mT the magnetisation deviates due to the formation of domains. The anisotropy field, $H_K$, was extrapolated from the coherent region to $M_z = 0$, $M_x = 1$, was found to be $\mu_0 H_K = 611$ mT. From unpatterned film SQUID magnetometry, we found the saturation magnetisation to be 0.83 MA m$^{-1}$, and noting the following relation

$$\mu_0 H_K = \frac{2 K_{eff}}{M_s} \quad (5)$$

where $M_s$ is the thin film saturation magnetisation, the effective anisotropy $K_{eff}$ is 0.51 MJ m$^{-3}$.

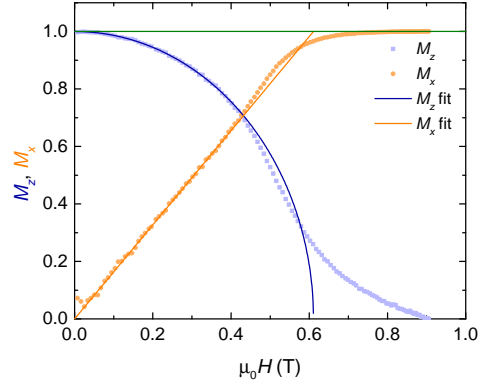

FIG. 1. Normalised AHE data for increasing in-plane applied magnetic field ($M_z$, square scatter plot), with its corresponding fit (solid line), and extrapolated in-plane magnetisation ($M_z$).

## COERCIVITY AND ANISOTROPY DISTRIBUTION

The anomalous Hall resistance of 8 non-irradiated samples was measured for $\mu_0 H_z$ and $\mu_0 H_x$ (Fig. 2). A significantly larger distribution of coercivities (14 ± 4 mT) was observed compared to the variation of anisotropy (1.32 ± 0.01 T).

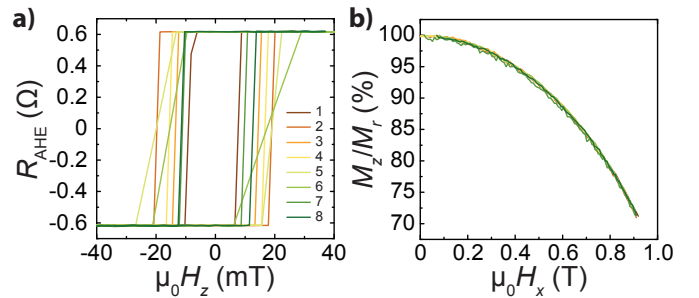

FIG. 2. Anomalous Hall resistance of virgin devices for **a)** $\mu_0 H_z$ and **b)** $\mu_0 H_x$. Panel b) shows the normalised first quadrant.

# THICKNESS-DEPENDENT CORRECTION OF RESISTIVITIES

The resistivity of metal thin films is known to increase significantly for thicknesses below 10 nm[2]. As the experimental resistivity is not always reported in the literature for SOT devices, we take bulk resistivity values and apply a consistent correction to the resistivity of each layer. Assuming grain-boundary dominated scattering, the thin film resistivity can be written using the M-S model as[3]:

$$\rho_f = \frac{\rho_0}{f_\alpha} \qquad (6)$$

where $\rho_f$ is the corrected thin film resistivity and $\rho_0$ is the bulk resistivity with

$$f_\alpha = 1 - \frac{3}{2}\alpha + 3\alpha^2 - 3\alpha^3 \ln\left(1 + \frac{1}{\alpha}\right) \qquad (7)$$

and

$$\alpha = \frac{\lambda R_{coef}}{D(1 - R_{coef})} \qquad (8)$$

where $\lambda$ is the electron mean free path without grain boundaries (typically 10 - 40 nm), $R_{coef}$ is the electron grain boundary reflection coefficient (typically $0.4 - 0.6$), and $D$ is the grain size. We fitted the M-S model to the measured resistivity of our complete stack, $\rho = 214$ $\mu\Omega.$cm, resulting in $\lambda = 25$ nm, $R_{coef} = 0.611$, and where we fix the grain size to the equal to the thickness of the layer, $D = d$. We then applied this model to determine the resistivity of the unreported literature stacks.

---